# A Nonparametric Statistical Method for Two Crossing Survival Curves[§]

Xinghui Huang [a], Jingjing Lyu [a], Yawen Hou [b], Zheng Chen [a*]

[a]Department of Biostatistics, Southern Medical University, Guangzhou, China.

[b]Department of Statistics, Jinan University, Guangzhou, China.

*Correspondence

Jan 2020

**Abstract**

In comparative research on time-to-event data for two groups, when two survival curves cross each other, it may be difficult to use the log-rank test and hazard ratio (HR) to properly assess the treatment benefit. Our aim was to identify a method for evaluating the treatment benefits for two groups in the above situation. We quantified treatment benefits based on an intuitive measure called the area between two survival curves (ABS), which is a robust measure of treatment benefits in clinical trials regardless of whether the proportional hazards assumption is violated or two survival curves cross each other. Additionally, we propose a permutation test based on the ABS, and we evaluate the effectiveness and reliability of this test with simulated data. The ABS permutation test is a robust statistical inference method with an acceptable type I error rate and superior power to detect differences in treatment effects, especially when the proportional hazards assumption is violated. The ABS can be used to intuitively quantify treatment differences over time and provide reliable conclusions in complicated situations, such as crossing survival curves.

The R Package "*ComparisonSurv*" contains the proposed methods and is available from https://CRAN.R-project.org/package=ComparisonSurv.

---

**Background**

Time-to-event end points are widely used in clinical follow-up studies. The difference in treatment benefits in survival analysis is commonly quantified and reported as the (cumulative) survival rate, the hazard ratio (HR), the median survival time (MT), or the restricted mean survival time (RMST) (Mantel, 1966; Pepe and Fleming, 1989; Royston and Parmar, 2011; Valérie and Sandrine et al., 2015; Rogawski and Westreich et al., 2016). The log-rank test corresponding to the HR is a common and classical choice when the HR between treatment arms remains constant over time (the proportional hazards assumption). However, when the proportional hazards assumption is violated, especially when two survival curves cross each other, the difference based on hazards may be offset before and after the crossing point, resulting in a loss of power of the log-rank test (Uno and Claggett et al., 2014; Li and Han et al., 2015). The method based on difference in the MT depends on information at the median survival point (when the survival rate is equal to 0.5) (Rogawski and Westreich et al., 2016; Zhao and Claggett et al., 2016). Therefore, when the censoring rate is large or the follow-up time is insufficient, this method cannot be applied without an accurate estimation of the MT (Saad and Zalcberg et al., 2018). The method based on the difference in RMST does not require the assumption of proportional hazards (Dehbi and Royston et al., 2017). However, when the two survival curves cross each other, the difference in the RMST may be offset before and after the crossing point, which causes an invalid assessment of the difference in treatment benefits and reduced power.

Hence, in this paper, we introduce a measure of the area between two survival curves (ABS) to interpret the difference in treatment benefits between two groups (Lin and Xu, 2010; Seruga and Pond et al., 2012; Zhao and Lu et al., 2012). Moreover, we propose a corresponding test that can avoid the problems stated above (Stefania and Maria Grazia, 2013). Simulations and examples are applied to compare the above measures. An R program has been developed to implement the ABS and the corresponding test.

**Methods**

The ABS is determined by the accumulated absolute difference in treatment effects from time 0 to the end of follow-up and can be estimated by integrating the absolute survival difference:

$$ABS = \int_0^v |S_1(t) - S_2(t)| dt \approx \sum_{j|t_j<v} |S_1(t_j) - S_2(t_j)|(t_{j+1} - t_j).$$

where $S_1(t), S_2(t)$ are the survival rates at time $t$ for each group, $t_j$ represents the pooled and ordered distinct event times in the two groups, and $v$ is the end of follow-up.

When comparing the differences between two groups, Lin and Xu (2010) estimated the mean $\hat{E}(ABS)$ and variance $\hat{V}(ABS)$ of the ABS under the null hypothesis and considered the normal transformation value $\Delta = \dfrac{ABS - \hat{E}(ABS)}{\sqrt{\hat{V}(ABS)}}$ to be a test statistic (ABS test) that would be an asymptotic standard normal distribution under the null hypothesis.

However, the test statistic $\Delta$ is non-normal in cases of small or large samples (details in Additional file 1). To provide a more robust confidence interval and a reliable hypothesis test result, we propose an ABS permutation test (ABSP test) based on permutation theory (Stefania and Maria Grazia, 2013). The ABSP test generates its own critical regions and $P$ value directly from the permutation resample data rather than from the distribution of the test statistic under the null hypothesis. The ABSP test procedure is as follows. First, $\Delta$ is computed from the original data; second, the ABS and the test statistic $\Delta(1)$ are computed for the 1st iteration. All observations are combined from both groups, and the original data are randomly resampled without replacement. Third, step 2 is repeated an additional 999 times, and $\Delta(2), \Delta(3), \Delta(4), \cdots, \Delta(1000)$ is computed. Finally, the two-sided $P$ value is computed based on the 1000 iterations: $P = \sum_{n=1}^{1000} I(n)/1000, n = 1, 2, \cdots 1000$, where $I(n)$ is an indicator function; $I(n)=1$ when $|\Delta(n)| > |\Delta|$ is satisfied and $I(n)=0$ otherwise.

Since researchers may also be interested in the short-term or long-term treatment effects, we extended the ABS to any time interval (ABSi) and created the ABS permutation over time interval test (ABSPi test). The null hypothesis is $H_0 : S_1(t) = S_2(t), t^* < t < t'', t^* \geq 0, t'' \leq v$, where $t^*$ and $t''$ are the left and right endpoints of the interval. The alternative hypothesis is $H_1 : S_1(t) \neq S_2(t)$ for some $t$ between $t^*$ and $t''$. Then, the ABS over the time interval is given by

$$ABS_{int} = \int_{t^*}^{t''} |S_1(t) - S_2(t)| dt \approx \sum_{j|t^* < t_j < t''} |S_1(t_j) - S_2(t_j)| (t_{j+1} - t_j).$$

After the endpoints of the interval are converted to $[0, t'']$ via an indicator function $\Phi(t_j > t^*)$, which is equal to 1 if $t_j > t^*$ and equal to 0 otherwise, $ABS_{int}$ can be expressed as

$$ABS_{int} = \int_{t^*}^{t''} |S_1(t) - S_2(t)| dt$$
$$= \int_0^{t''} \Phi(t > t^*) |S_1(t) - S_2(t)| dt \approx \sum_{j|t_j < t''} |S_1(t_j) - S_2(t_j)| [\Phi(t_j > t^*)(t_{j+1} - t_j)].$$

Therefore, based on the Kaplan-Meier estimation, we have

$$ABS_{int} \approx \sum_{j|t_j < t''} |\hat{S}_1(t_j) - \hat{S}_2(t_j)| [\Phi(t_j > t^*)(t_{j+1} - t_j)].$$

Under the null hypothesis, $\hat{S}_1(t) - \hat{S}_2(t)$ follows an approximately normal distribution with a mean of 0 and variance of $[\hat{\sigma}_{S_1}^2(t) + \hat{\sigma}_{S_2}^2(t)]$. Then, we have the following:

$$\hat{E}(|\hat{S}_1(t) - \hat{S}_2(t)|) = \{\frac{2}{\pi} [\hat{\sigma}_{S_1}^2(t) + \hat{\sigma}_{S_2}^2(t)]\}^{1/2},$$

$$\hat{V}\left(\left|\hat{S}_1(t) - \hat{S}_2(t)\right|\right) = (1 - \frac{2}{\pi})[\hat{\sigma}_{S_1}^2(t) + \hat{\sigma}_{S_2}^2(t)].$$

Based on the normal approximation, we can estimate the mean and variance of $ABS_{int}$ as follows:

$$\hat{E}(ABS_{int}) = \sum_{j|t_j<t''} \left\{\frac{2}{\pi}[\hat{\sigma}_{S_1}^2(t_j) + \hat{\sigma}_{S_2}^2(t_j)]\right\}^{1/2} \Phi(t_j > t^*)(t_{j+1} - t_j),$$

$$\hat{V}(ABS_{int}) = \sum_{j|t_j<t''} \Phi(t_j > t^*)(t_{j+1} - t_j)^2 (1 - \frac{2}{\pi})\left(\hat{\sigma}_{S_1}^2(t_j) + \hat{\sigma}_{S_2}^2(t_j)\right) +$$

$$\sum_{j<j'|t_j,t_{j'}<t''} 2\rho_{j,j'} \Phi(t_j > t^*)\Phi(t_{j'} > t^*)(t_{j+1}-t_j)(t_{j'+1}-t_{j'})(1-\frac{2}{\pi}) \times \left\{[\hat{\sigma}_{S_1}^2(t_j) + \hat{\sigma}_{S_2}^2(t_j)][\hat{\sigma}_{S_1}^2(t_{j'}) + \hat{\sigma}_{S_2}^2(t_{j'})]\right\}^{1/2}.$$

where $\rho_{j,j'}$ is the correlation coefficient between $\left|\hat{S}_1(t_j) - \hat{S}_2(t_j)\right|$ and $\left|\hat{S}_1(t_{j'}) - \hat{S}_2(t_{j'})\right|$ and $j < j'$. Lin (2010) suggested setting $\rho_{j,j'}$ to 0.5 for all $j$ and $j'$. Therefore, we have the test statistic $\Delta_{int} = [ABS_{int} - \hat{E}(ABS_{int})] / \sqrt{\hat{V}(ABS_{int})}$. Using permutation theory, we can compute the *P* value based on 1000 resamples.

## Simulation study

### Simulation design

To evaluate and compare the statistical performance of the ABSP test with that of the log-rank test, the RMST test, and the ABS test, we calculated the type I error rates and power via Monte Carlo simulations (the test based on the difference in MT is excluded here because the survival rate of any group may not reach 0.5, which cannot yield a test result). We considered the following situations (shown in Figure 1): Situation I. two survival curves overlapping (the differences between two groups are generated by random sampling to evaluate false positive rates of tests under the null hypothesis, that is, type I error rates); Situation II. two survival curves with proportional hazard functions; Situation III. two survival curves with early and middle differences; Situation IV. two survival curves with a late difference; Situation V. two survival curves with median crossing; and Situation VI. two survival curves with late crossing.

We considered censoring rates of 0%, 15%, 30%, and 45% as well as groups of equal sample sizes ($N_1=N_2=20, 50, 100$) or unequal sample sizes ($N_1=20, N_2=50$; $N_1=50, N_2=100$). The lifetime *X* was generated from an exponential or Weibull distribution, and the variable *status* was set as $\delta = 1$ when there was no censoring. When we considered right-censored data, we assumed that patients entered the study uniformly over a time interval and that censoring time *C* was generated from uniform distributions *U* (0, *a*) and *U* (0, *b*). We denoted the lifetime $T = \min(X, C)$, $\delta = I[X \leq C]$, where *I* (·) is an indicator function. We could adjust the average censoring rates through the values of *a* and *b*. The number of iterations in each simulation was 1000, and we would reject the null hypothesis at the 0.05 significance level.

**Simulation results**

To present the simulation results clearly, we provide the summarized results of the type I error rates (Table 1) and power for several tests (Table 2), where the summarized results are obtained by variance (ANOVA) techniques (original results in Additional file 2) (Logan and Klein et al., 2008; Klein and Brent et al., 2010). To evaluate the type 1 error rates, the response variable *Y* was defined as the percent rejection rate minus the nominal 5% level; therefore, good performance is implied by close-to-zero estimates for the expectation E(*Y*) in the ANOVA. To evaluate power, the outcome variable *Y* was defined as the rejection rate: good performance is indicated by large estimated values of E(*Y*). Here, we consider four factors: *TEST*, with 4 levels; *NUM*, with 5 levels; *CENSORE*, with 4 levels; and *SITUATION*, with 6 levels. These factors represent the test method, sample size of each group, censoring rate, and simulation situation, respectively. We fit four models without intercepts for E(*Y*):

Model 1:  $E(Y) = TEST \times NUM + CENSORE + SITUATION$.

Model 2:  $E(Y) = TEST \times CENSORE + NUM + SITUATION$.

Model 3: $E(Y) = TEST \times SITUATION + CENSORE + NUM$.

Model 4: $E(Y) = TEST + SITUATION + CENSORE + NUM$.

As shown in Table 1, the type I error rates of the ABS test deviate from the prespecified nominal level, whereas those of the other tests are close to 0.05. The censoring rates and sample sizes have little impact on the type I error rate.

As shown in Table 2, in situation II, the four tests have similar mean rejection rates, and the log-rank test performs best. The ABSP test is robust and has the highest mean rejection rates in other situations. Whether the sample sizes of the two groups are equal has little effect on the power.

The log-rank test may have worse power when the proportional hazards assumption is violated. Tests based on a difference in RMST may lose power when the survival curves cross because of differences in offsetting before and after the crossing point. The ABS test is not reliable because the type I error rates deviate from the prespecified nominal level. The ABSP test is a robust test for comparing treatment benefits, regardless of whether the proportional hazards assumption is violated or the survival curves cross.

**Examples**

Example 1: In a clinical trial on the role of docetaxel in non-metastatic castration-resistant prostate cancer (nmCRPC) (Ito and Kimura et al., 2018), 46 patients received docetaxel for nmCRPC, and 52 had distant metastases. As shown in Figure 2A (simulated data from Figure 3A in the original paper), the Kaplan-Meier curves of overall survival in the two groups are separate, and the proportional hazards assumption is satisfied ($P$=0.881). As shown in Example 1 in Table 3, the test based on the difference in the MT cannot provide a result because the survival rate of the nmCRPC group did not reach 0.5, whereas the other tests indicate that the treatment effect in the nmCRPC group was better than that in the placebo group. Thus, the treatment effect in the nmCRPC group was better than that in the mCRPC group in both the short term and the long term (Example 1 in Table 3).

Example 2: The kidney-dialysis trial was designed to assess the time to the first exit-site infection (in months) in patients with renal insufficiency (Yan, 2004). In 43 patients, the catheter was surgically implanted, whereas in 76 patients, the catheter was percutaneously placed. There were 27 patients censored in the first group and 65 in the second group. As shown in Figure 2B, the two survival curves cross at an early time, and the Grambsch-Therneau test indicates that the proportional hazards assumption is violated ($P$=0.003). As shown in Example 2 in Table 3, the test based on the difference in the MT cannot provide a result as in Example 1, and the log-rank test indicates no significant difference, whereas the test based on the difference in the RMST and the ABSP test yield significant overall results. Further, the treatment effect in the percutaneously placed catheter group is better than that in the surgically implanted catheter group in the long term (after 8 months), which may result in an overall significant difference.

Example 3: In a randomized clinical trial on arginine deprivation with pegylated arginine deiminase in patients with argininosuccinate synthetase 1-deficient malignant pleural mesothelioma (Szlosarek and Steele et al., 2016), 68 patients were randomized to receive the arginine-lowering agent pegylated arginine deiminase (ADI-PEG20, 36.8 mg/m$^2$, weekly, intramuscular) plus best supportive care (Group 1) or best supportive care alone (Group 2). As shown in Figure 2C (simulated data from Figure 2B in the original paper with approximately the same sample size, censoring rate, and basic statistics; the sample sizes of Group 1 and Group 2 were 44 and 24 patients, and the censoring rates in the two groups were approximately 4.5% and 4.2%, respectively), the two overall survival curves cross in the 10th month, and the survival rates of the two groups reverse before and after the crossing point, which may be caused by the treatment effects between the two groups differing from the early to late phases. The Grambsch-Therneau test indicates that the proportional hazards assumption is violated ($P$=0.029).

The descriptive statistics and test results based on the HR, MT, RMST, and ABS are shown in Table 3. The HR of Group 1 is 0.673 times that of Group 2 (HR=0.673; 95% CI:

0.394 to 1.148; *P*=0.147); the difference in the MT between groups is 0.120 (95% CI: -0.064 to 0.308; *P*>0.999) months; and the difference in RMST between the two groups is 3.714 (95% CI: -1.126 to 8.553; *P*=0.133) months. These three measures indicate no statistical significance for the overall test. However, the ABS, which reflects the absolute survival difference between the two groups, is 5.440 (95% CI: 2.914 to 8.888) months, indicating a significant difference (ABSP *P*=0.031). Since the two survival curves cross in the 10th month, which indicates that the treatment effects of the two groups may reverse before and after the crossing point, we used the ABSPi test to analyze the interval differences before and after the crossing point. As shown in Table 3, the ABSPi test indicates that the short-term treatment effects of the two groups are not significantly different before the 10th month from the start of follow-up (*P*=0.360), whereas the differences in long-term treatment effects of the two groups are statistically significant. The treatment effect of Group 1 is better than that of Group 2 at the 10th month or later (*P*=0.028). In other words, no difference between treatments is observed in the short term, while the treatment effect in Group 1 is better than that in Group 2 in the long term, resulting in an overall significant difference (ABSP *P*=0.031).

**Discussion**

When the proportional hazards assumption is violated, especially when two survival curves cross, the HR may change over time; thus, a constant HR cannot be used to interpret the differences in overall survival. Therefore, the corresponding log-rank test is not appropriate. A test based on a difference in MT, which does not require the proportional hazards assumption and uses information only at the MT, may lose information and may be affected by the scheduled endpoint assessment rather than overall survival (Hajime and Brian et al., 2014; Rogawski and Westreich et al., 2016; Saad and Zalcberg et al., 2018). Meanwhile, a test based on a difference in MT is not applicable when the survival rate does not reach 0.5 or when the crossing point of the survival curves is located near 0.5. A test based on the difference in the RMST is easily calculated and interpreted according to the difference in areas under survival curves (Royston and Parmar, 2011). This test is typically not affected by the proportional hazards assumption; however, when the survival curves cross, as in Example 2 and Example 3 (Figures 2B/2C), the RMSTs of Group 1 and Group 2 can be expressed as "area $D_2$+S" and "area $D_1$+S". Thus, the difference in the RMST can be estimated as "$D_2$-$D_1$", which is the same as the differences in the ABS before and after the crossing point. Therefore, a test based on the difference in the RMST gives only the relative difference in effects and does not lose power in this case because of differences in offsetting before and after the crossing point (the same occurs in Situations V and VI). However, the ABS can be expressed as "area $D_1$+$D_2$", which reflects the absolute difference in treatment effects from the start to the end of follow-up. The ABS is not affected by violation of the proportional hazards assumption or crossing of the survival curves. When the survival curves are separated, a test based on the ABS yields a similar result to the test based on a difference in the RMST.

**Conclusions**

The ABS measure effectively reflects the absolute benefit of treatment effects between two groups in randomized clinical trials. The ABS is a robust measure for comparing the treatment benefits of two groups and can be applied regardless of whether the proportional hazards assumption is violated or the survival curves cross. Meanwhile, the corresponding ABSP test is also robust and reliable. When researchers are interested in treatment effects over any time interval (i.e., evaluation of the short-term or long-term treatment benefits of two cancer drugs), the corresponding ABSPi test can be applied in combination with Kaplan-Meier curves to draw a comprehensive conclusion. Additionally, an R program was developed for the application of the ABS and the corresponding test (details in Additional file 3).

**Abbreviations**

HR: hazard ratio; MT: median survival time; RMST: restricted mean survival time; ABS: area between two survival curves; $ABSi_L$/$ABSi_R$: area between two survival curves over the left/right time interval; ABSP: area between two survival curves permutation test; MTd: difference in median survival time; RMSTd: difference in restricted mean survival time.

**Competing interests**

The authors declare that they have no competing interests.

**Funding**

This work is supported by the National Natural Science Foundation of China (81673268, 81903411) and Natural Science Foundation of Guangdong Province (2017A030313812, 2018A030313849).

**Acknowledgements**

Not applicable.

**Table 1.** Average deviations from the nominal 5 percent level (Type I error rates).

|  | TEST | Log-rank | RMSTd | ABS | ABSP |
|---|---|---|---|---|---|
| *NUM* * | (20, 20) | -0.650 | -1.450 | -2.300 | -1.250 |
|  | (50, 50) | -0.425 | -0.050 | -1.825 | 0.050 |
|  | (100, 100) | 0.250 | -0.300 | -2.500 | -0.250 |
|  | (20, 50) | -0.125 | 0.000 | -1.000 | -0.250 |
|  | (50, 100) | -0.625 | -0.475 | -2.150 | -0.575 |
| *CENSORE* ¶ | 0.00 | -0.020 | -0.960 | -2.820 | -0.540 |
|  | 0.15 | -0.020 | -0.940 | -2.580 | -0.500 |
|  | 0.30 | -0.480 | 0.040 | -1.520 | -0.280 |
|  | 0.45 | -0.060 | 0.040 | -0.900 | -0.500 |
| *SITUATION* ⨆ | I | -0.145 | -0.455 | -1.955 | -0.455 |
| TOTAL§ |  | -0.145 | -0.455 | -1.955 | -0.455 |

*: Model 1; ¶: Model 2; ⨆: Model 3; §: Model 4; *CENSORE*: censoring rate;
RMSTd: test based on the difference in the RMST; ABS: ABS test; ABSP: ABS permutation test;
default range of average deviations for 1000 iterations: [-1.351, 1.351].

**Table 2.** Average rejection rates (power).

|  | TEST | Log-rank | RMSTd | ABS | ABSP |
|---|---|---|---|---|---|
| *NUM* * | (20, 20) | 0.233 | 0.240 | 0.253 | 0.308 |
|  | (50, 50) | 0.400 | 0.408 | 0.562 | 0.643 |
|  | (100, 100) | 0.523 | 0.558 | 0.782 | 0.836 |
|  | (20, 50) | 0.283 | 0.289 | 0.447 | 0.464 |
|  | (50, 100) | 0.431 | 0.467 | 0.680 | 0.719 |
| *CENSORE* ¶ | 0.00 | 0.432 | 0.425 | 0.669 | 0.757 |
|  | 0.15 | 0.366 | 0.371 | 0.610 | 0.680 |
|  | 0.30 | 0.326 | 0.342 | 0.534 | 0.566 |
|  | 0.45 | 0.372 | 0.432 | 0.366 | 0.373 |
| *SITUATION* ⨆ | II | **0.874** | 0.836 | 0.746 | 0.768 |
|  | III | 0.195 | 0.296 | 0.275 | **0.361** |
|  | VI | 0.414 | 0.301 | 0.590 | **0.636** |
|  | V | 0.192 | 0.216 | 0.501 | **0.539** |
|  | VI | 0.196 | 0.314 | 0.611 | **0.667** |
| TOTAL§ |  | 0.374 | 0.392 | 0.545 | **0.594** |

*: Model 1; ¶: Model 2; ⨆: Model 3; §: Model 4; *CENSORE*: censoring rate;
RMSTd: test based on the difference in the RMST; ABS: ABS test; ABSP: ABS permutation test.

**Table 3.** Results of different measures of the treatment effect for three examples.

| Measure | Example 1 | | Example 2 | | Example 3 | |
|---|---|---|---|---|---|---|
| | Statistic (95% CI) | $P$ | Statistic (95% CI) | $P$ | Statistic (95% CI) | $P$ |
| HR | 0.3789 (0.196, 0.732) | $0.003^a$ | 0.542 (0.249, 1.182) | $0.118^a$ | 0.673 (0.394, 1.148) | $0.147^a$ |
| MTd | / | / | / | / | 0.120 (-0.064, 0.308) | 0.999 |
| RMSTd | 38.943 (15.414, 62.472) | 0.001 | 4.438 (0.185, 8.491) | 0.041 | 3.714 (-1.126, 8.553) | 0.133 |
| ABS | 38.943 | $0.003^b$ | 5.120 | $0.030^b$ | 5.440 | $0.031^b$ |
| $ABSi_L$ | 14.039 | $0.009^c$ | 0.378 | $0.805^c$ | 0.858 | $0.360^c$ |
| $ABSi_R$ | 24.904 | $0.004^c$ | 4.473 | $0.024^c$ | 4.581 | $0.028^c$ |

95% CI: 95% confidence interval; $^a$: log-rank test; $^b$: ABSP test; $^c$: ABSPi test; $ABSi_L/ABSi_R$: ABS over the left/right time interval, [0, 75] and [75,156.383] for Example 1, [0, 8] and [8, 28.5] for Example 2, [0, 10] and [10, 39.4] for Example 3; MTd/RMSTd: difference in the MT/RMST.

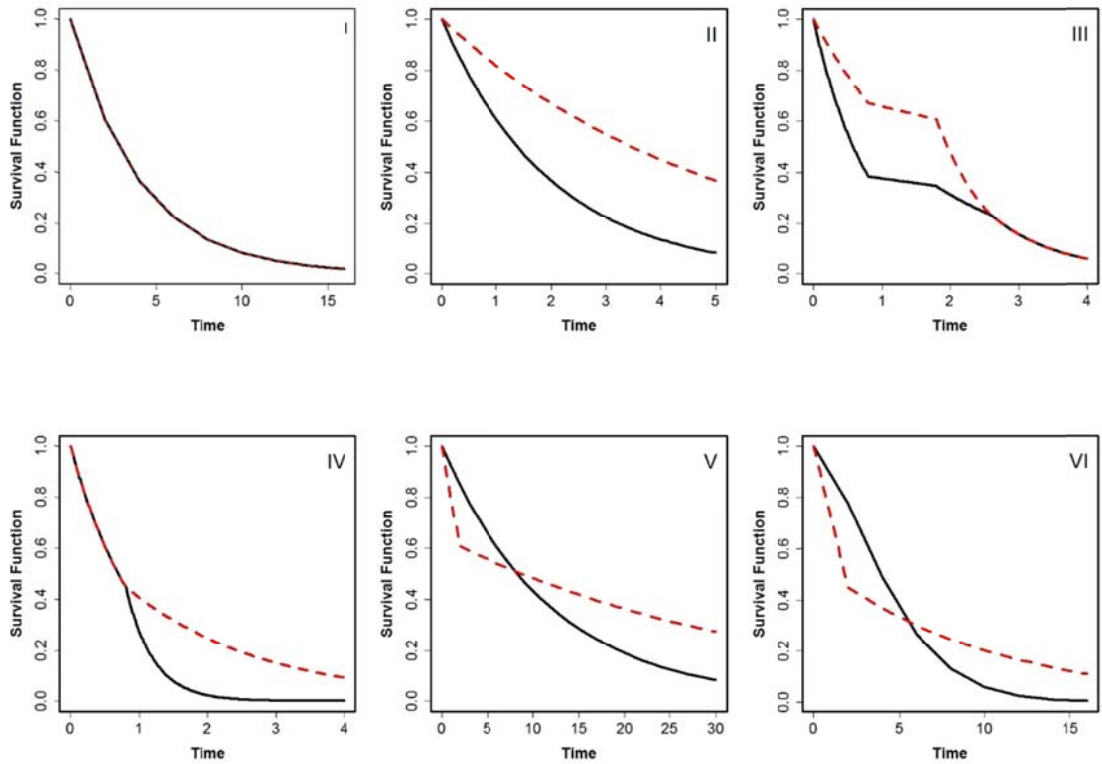

**Figure 1:** Plot of survival curves for the six simulation situations. The survival times are from a piecewise exponential distribution with hazard rate $\lambda_i(t)$ or Weibull distribution ($W(\eta,\theta)$, where $\eta>0$ is the shape parameter and $\theta>0$ is the scale parameter) for groups 1 and 2. Specifically, Situation I. $\lambda_1=\lambda_2=0.25$. Situation II. $\lambda_1=0.5$ and $\lambda_2=0.2$. Situation III. $\lambda_1=1.2,0.1,0.5,1$ and $\lambda_2=0.5,0.1,1.2,1$ for $t\le 0.8, 0.8<t\le 1.8, 1.8<t\le 2.6, t>2.6$, respectively. Situation IV. $\lambda_1=1,2.5$ and $\lambda_2=1,0.5$ for $t\le 0.8, t>0.8$, respectively. Situation V. $\lambda_1=1/12$ and $\lambda_2=1/4,1/35$ for $t\le 2, t>2$, respectively. Situation VI. $W(1.5,5)$ for group 1; $\lambda=0.5,0.1$ for $t\le 1.5, t>1.5$, respectively, for group 2.

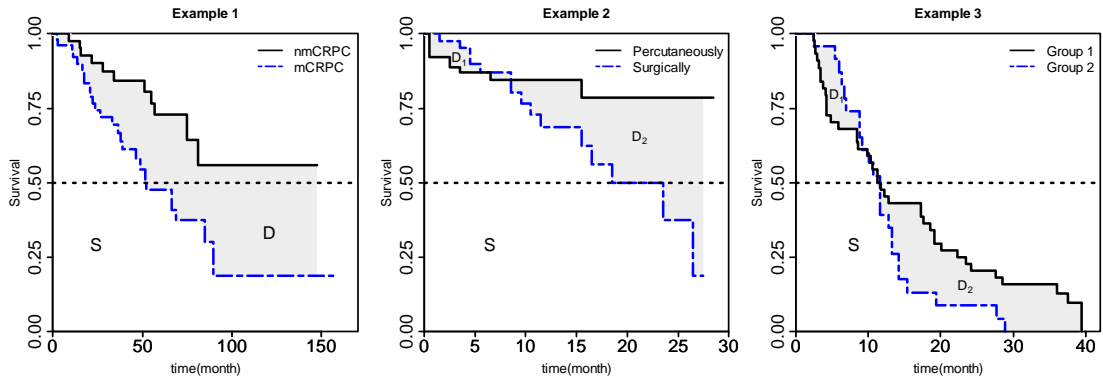

**Figure 2:** Kaplan-Meier curves for the three examples. D in Example 1: ABS; S: shared area under the survival curves of the two groups; $D_1$, $D_2$ in Example 2/3: ABS before and after the crossing point of the survival curves.

**Additional file 1: Study on the normality of the test statistic.**

Assume a randomized study of two groups with sample size $n_k$, where $k = 1, 2$. $t_{ki}$ denotes the survival time, and $\delta_{ki}$ denotes the status of the $i$th case in group $k$, where $\delta_{ki}=1$ indicates an observed event, and $\delta_{ki}=0$ indicates censored status. Let $t_1 < t_2 < \cdots < t_m$ represent the pooled distinct event times in the two samples. $d_{kj}$ denotes the number of observed events, and $Y_{kj}$ denotes the number of subjects at risk at time point $t_j$, where $j = 1, \cdots, m$. Then, the survival rate and its variance can be expressed as: $\hat{S}_k(t) = \prod_{j|t_j \leq t}(1 - d_{kj}/Y_{kj})$ and

$$\hat{\sigma}^2_{S_k}(t) = \hat{S}_k(t)^2 \sum_{j|t_j \leq t} \frac{d_{kj}}{Y_{kj}(Y_{kj} - d_{kj})}.$$

The measure of the area between two survival curves (ABS) can be given by:

$$ABS = \int_0^v |S_1(t) - S_2(t)| dt \approx \sum_{j|t_j < v} |S_1(t_j) - S_2(t_j)|(t_{j+1} - t_j)$$

where $S_1(t), S_2(t)$ are the survival rates at time $t$ for each group, and $v$ is the last time point for which the areas under the survival curves can be calculated for both groups based on the available data. Specifically, $v = \min(t_{1n_1}, t_{2n_2})$ if the last time points in the two groups are both censored; $v = \max[t_{1n_1}(1-\delta_{1n_1}), t_{2n_2}(1-\delta_{2n_2})]$ if the last time point in one group is an actual event and that in the other group is censored; and $v = \max(t_{1n_1}, t_{2n_2})$ if the last time points in both groups are actual events. Denote $\Delta = \dfrac{ABS - \hat{E}(ABS)}{\sqrt{\hat{V}(ABS)}}$ by normal transformation, which will be an asymptotic standard normal distribution.

We first calculate the value of $\Delta$ under the null hypothesis and then plot the corresponding density histograms and fitted curves. The Shapiro-Wilk normality test is also used to investigate the asymptotical normality of the test statistic. We consider censoring rates (CRs) of 0%, 15%,

30%, and 45% and groups of equal sample size ($N_1=N_2=20, 50, 100$) or unequal sample sizes ($N_1=20, N_2=50$; $N_1=50, N_2=100$). As shown in eFigure 1 (the horizontal axis denotes the value of $\Delta$, the vertical axis denotes the density), the test statistic $\Delta$ is nonnormal (right trailing approximatively) for different sample sizes and CRs. Meanwhile, the results of the Shapiro-Wilk normality test show that the test statistic $\Delta$ is not normally distributed ($P<0.001$).

**eFigure 1.** Density histogram and fitted line of the test statistic ($\Delta$) for different sample sizes and censoring rates.

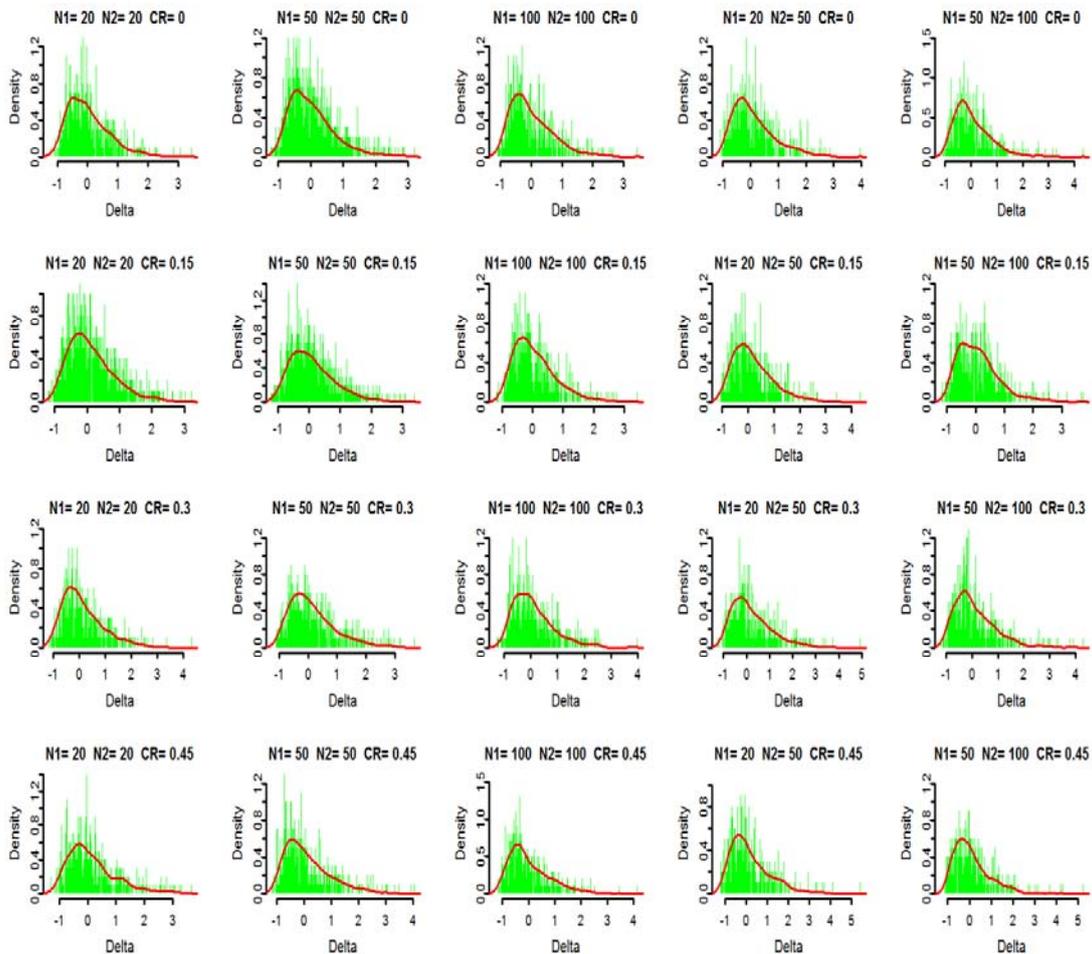

# Additional file 2: Original results for simulations

**eTable 1. Type I error rates (I).**

| N1 | N2 | Censore | Log-rank | RMSTd | ABS | ABSP |
|---|---|---|---|---|---|---|
| 20 | 20 | 0.00 | 0.042 | 0.031 | 0.013 | 0.040 |
|    |    | 0.15 | 0.040 | 0.031 | 0.021 | 0.036 |
|    |    | 0.30 | 0.040 | 0.040 | 0.029 | 0.035 |
|    |    | 0.45 | 0.052 | 0.040 | 0.045 | 0.039 |
| 50 | 50 | 0.00 | 0.056 | 0.050 | 0.025 | 0.044 |
|    |    | 0.15 | 0.059 | 0.040 | 0.023 | 0.049 |
|    |    | 0.30 | 0.053 | 0.057 | 0.035 | 0.057 |
|    |    | 0.45 | 0.049 | 0.053 | 0.044 | 0.052 |
| 100 | 100 | 0.00 | 0.059 | 0.040 | 0.021 | 0.045 |
|    |    | 0.15 | 0.048 | 0.043 | 0.018 | 0.049 |
|    |    | 0.30 | 0.041 | 0.059 | 0.030 | 0.049 |
|    |    | 0.45 | 0.062 | 0.056 | 0.031 | 0.047 |
| 20 | 50 | 0.00 | 0.050 | 0.039 | 0.028 | 0.054 |
|    |    | 0.15 | 0.053 | 0.042 | 0.033 | 0.047 |
|    |    | 0.30 | 0.050 | 0.045 | 0.050 | 0.048 |
|    |    | 0.45 | 0.042 | 0.055 | 0.049 | 0.041 |
| 50 | 100 | 0.00 | 0.042 | 0.042 | 0.022 | 0.040 |
|    |    | 0.15 | 0.049 | 0.047 | 0.026 | 0.044 |
|    |    | 0.30 | 0.042 | 0.051 | 0.030 | 0.047 |
|    |    | 0.45 | 0.042 | 0.048 | 0.036 | 0.046 |

Censore: censoring rate; RMSTd: test based on RMSTd; ABS: ABS test; ABSP: ABS Permutation test;

Default range for 1000 iterations: [0.036, 0.064].

**eTable 2. Power for the scenario of curves with proportional hazards (II).**

| N1 | N2 | Censore | Log-rank | RMSTd | ABS | ABSP |
|----|----|---------|----------|-------|-----|------|
| 20 | 20 | 0.00 | 0.771 | 0.686 | 0.623 | 0.745 |
|    |    | 0.15 | 0.685 | 0.603 | 0.535 | 0.619 |
|    |    | 0.30 | 0.596 | 0.532 | 0.478 | 0.490 |
|    |    | 0.45 | 0.472 | 0.422 | 0.398 | 0.349 |
| 50 | 50 | 0.00 | 0.995 | 0.870 | 0.968 | 0.993 |
|    |    | 0.15 | 0.987 | 0.788 | 0.908 | 0.947 |
|    |    | 0.30 | 0.954 | 0.685 | 0.748 | 0.791 |
|    |    | 0.45 | 0.865 | 0.526 | 0.620 | 0.621 |
| 100 | 100 | 0.00 | 1.000 | 0.988 | 1.000 | 1.000 |
|    |    | 0.15 | 1.000 | 0.973 | 0.987 | 0.994 |
|    |    | 0.30 | 1.000 | 0.931 | 0.903 | 0.937 |
|    |    | 0.45 | 0.988 | 0.836 | 0.748 | 0.782 |
| 20 | 50 | 0.00 | 0.914 | 1.000 | 0.889 | 0.925 |
|    |    | 0.15 | 0.861 | 0.994 | 0.727 | 0.764 |
|    |    | 0.30 | 0.797 | 0.980 | 0.607 | 0.600 |
|    |    | 0.45 | 0.667 | 0.915 | 0.476 | 0.428 |
| 50 | 100 | 0.00 | 1.000 | 1.000 | 0.999 | 0.999 |
|    |    | 0.15 | 0.998 | 1.000 | 0.926 | 0.955 |
|    |    | 0.30 | 0.987 | 0.999 | 0.774 | 0.802 |
|    |    | 0.45 | 0.940 | 0.987 | 0.609 | 0.622 |

Censore: censoring rate; RMSTd: test based on RMSTd; ABS: ABS test; ABSP: ABS Permutation test.

**eTable 3. Power for the scenario of curves with early and middle differences (III).**

| N1 | N2 | Censore | Log-rank | RMSTd | ABS | ABSP |
|---|---|---|---|---|---|---|
| 20 | 20 | 0.00 | 0.072 | 0.128 | 0.101 | 0.143 |
| | | 0.15 | 0.086 | 0.131 | 0.110 | 0.142 |
| | | 0.30 | 0.111 | 0.172 | 0.119 | 0.133 |
| | | 0.45 | 0.152 | 0.225 | 0.148 | 0.138 |
| 50 | 50 | 0.00 | 0.089 | 0.164 | 0.251 | 0.393 |
| | | 0.15 | 0.123 | 0.172 | 0.218 | 0.354 |
| | | 0.30 | 0.189 | 0.222 | 0.246 | 0.346 |
| | | 0.45 | 0.314 | 0.278 | 0.286 | 0.334 |
| 100 | 100 | 0.00 | 0.117 | 0.240 | 0.533 | 0.720 |
| | | 0.15 | 0.186 | 0.225 | 0.487 | 0.690 |
| | | 0.30 | 0.295 | 0.270 | 0.503 | 0.653 |
| | | 0.45 | 0.511 | 0.473 | 0.507 | 0.579 |
| 20 | 50 | 0.00 | 0.120 | 0.284 | 0.163 | 0.213 |
| | | 0.15 | 0.135 | 0.264 | 0.163 | 0.210 |
| | | 0.30 | 0.193 | 0.305 | 0.175 | 0.217 |
| | | 0.45 | 0.265 | 0.523 | 0.199 | 0.198 |
| 50 | 100 | 0.00 | 0.125 | 0.376 | 0.338 | 0.483 |
| | | 0.15 | 0.173 | 0.360 | 0.300 | 0.456 |
| | | 0.30 | 0.239 | 0.421 | 0.313 | 0.424 |
| | | 0.45 | 0.395 | 0.682 | 0.340 | 0.386 |

Censore: censoring rate; RMSTd: test based on RMSTd; ABS: ABS test; ABSP: ABS Permutation test.

**eTable 4. Power for the scenario of curves with a late difference (IV).**

| N1  | N2  | Censore | Log-rank | RMSTd | ABS   | ABSP  |
|-----|-----|---------|----------|-------|-------|-------|
| 20  | 20  | 0.00    | 0.333    | 0.138 | 0.283 | 0.467 |
|     |     | 0.15    | 0.210    | 0.099 | 0.243 | 0.339 |
|     |     | 0.30    | 0.120    | 0.083 | 0.235 | 0.238 |
|     |     | 0.45    | 0.082    | 0.070 | 0.147 | 0.129 |
| 50  | 50  | 0.00    | 0.756    | 0.223 | 0.845 | 0.965 |
|     |     | 0.15    | 0.577    | 0.136 | 0.741 | 0.875 |
|     |     | 0.30    | 0.332    | 0.085 | 0.584 | 0.657 |
|     |     | 0.45    | 0.144    | 0.060 | 0.283 | 0.304 |
| 100 | 100 | 0.00    | 0.962    | 0.523 | 0.998 | 1.000 |
|     |     | 0.15    | 0.882    | 0.390 | 0.985 | 0.993 |
|     |     | 0.30    | 0.649    | 0.186 | 0.882 | 0.923 |
|     |     | 0.45    | 0.229    | 0.073 | 0.392 | 0.471 |
| 20  | 50  | 0.00    | 0.417    | 0.701 | 0.722 | 0.827 |
|     |     | 0.15    | 0.261    | 0.527 | 0.615 | 0.663 |
|     |     | 0.30    | 0.169    | 0.274 | 0.512 | 0.490 |
|     |     | 0.45    | 0.092    | 0.097 | 0.284 | 0.231 |
| 50  | 100 | 0.00    | 0.862    | 0.899 | 0.991 | 0.997 |
|     |     | 0.15    | 0.653    | 0.790 | 0.929 | 0.962 |
|     |     | 0.30    | 0.394    | 0.517 | 0.748 | 0.785 |
|     |     | 0.45    | 0.151    | 0.145 | 0.388 | 0.405 |

Censore: censoring rate; RMSTd: test based on RMSTd; ABS: ABS test; ABSP: ABS Permutation test.

**eTable 5. Power for the scenario of curves with median crossing (V).**

| N1 | N2 | Censore | Log-rank | RMSTd | ABS | ABSP |
|---|---|---|---|---|---|---|
| 20 | 20 | 0.00 | 0.205 | 0.117 | 0.227 | 0.388 |
| | | 0.15 | 0.107 | 0.068 | 0.204 | 0.275 |
| | | 0.30 | 0.055 | 0.053 | 0.200 | 0.183 |
| | | 0.45 | 0.053 | 0.127 | 0.118 | 0.094 |
| 50 | 50 | 0.00 | 0.422 | 0.170 | 0.688 | 0.863 |
| | | 0.15 | 0.230 | 0.091 | 0.585 | 0.709 |
| | | 0.30 | 0.087 | 0.051 | 0.445 | 0.475 |
| | | 0.45 | 0.055 | 0.093 | 0.210 | 0.206 |
| 100 | 100 | 0.00 | 0.759 | 0.382 | 0.990 | 0.998 |
| | | 0.15 | 0.449 | 0.228 | 0.920 | 0.962 |
| | | 0.30 | 0.162 | 0.083 | 0.787 | 0.829 |
| | | 0.45 | 0.055 | 0.077 | 0.468 | 0.516 |
| 20 | 50 | 0.00 | 0.178 | 0.586 | 0.550 | 0.642 |
| | | 0.15 | 0.098 | 0.352 | 0.442 | 0.457 |
| | | 0.30 | 0.045 | 0.120 | 0.350 | 0.300 |
| | | 0.45 | 0.033 | 0.079 | 0.186 | 0.132 |
| 50 | 100 | 0.00 | 0.489 | 0.797 | 0.947 | 0.977 |
| | | 0.15 | 0.231 | 0.585 | 0.807 | 0.862 |
| | | 0.30 | 0.075 | 0.201 | 0.596 | 0.607 |
| | | 0.45 | 0.045 | 0.060 | 0.301 | 0.299 |

Censore: censoring rate; RMSTd: test based on RMSTd; ABS: ABS test; ABSP: ABS Permutation test.

**eTable 6. Power for the scenario of curves with late crossing (VI).**

| N1 | N2 | Censore | Log-rank | RMSTd | ABS | ABSP |
|---|---|---|---|---|---|---|
| 20 | 20 | 0.00 | 0.048 | 0.083 | 0.191 | 0.450 |
|  |  | 0.15 | 0.071 | 0.132 | 0.207 | 0.337 |
|  |  | 0.30 | 0.107 | 0.293 | 0.254 | 0.280 |
|  |  | 0.45 | 0.318 | 0.639 | 0.241 | 0.230 |
| 50 | 50 | 0.00 | 0.060 | 0.070 | 0.827 | 0.983 |
|  |  | 0.15 | 0.053 | 0.107 | 0.766 | 0.909 |
|  |  | 0.30 | 0.154 | 0.262 | 0.648 | 0.734 |
|  |  | 0.45 | 0.607 | 0.720 | 0.364 | 0.409 |
| 100 | 100 | 0.00 | 0.070 | 0.073 | 1.000 | 1.000 |
|  |  | 0.15 | 0.048 | 0.098 | 0.994 | 1.000 |
|  |  | 0.30 | 0.228 | 0.265 | 0.966 | 0.988 |
|  |  | 0.45 | 0.863 | 0.836 | 0.588 | 0.676 |
| 20 | 50 | 0.00 | 0.012 | 0.064 | 0.615 | 0.750 |
|  |  | 0.15 | 0.024 | 0.096 | 0.514 | 0.568 |
|  |  | 0.30 | 0.062 | 0.281 | 0.416 | 0.384 |
|  |  | 0.45 | 0.316 | 0.905 | 0.338 | 0.280 |
| 50 | 100 | 0.00 | 0.012 | 0.058 | 0.993 | 0.998 |
|  |  | 0.15 | 0.030 | 0.070 | 0.932 | 0.969 |
|  |  | 0.30 | 0.146 | 0.274 | 0.851 | 0.882 |
|  |  | 0.45 | 0.684 | 0.947 | 0.513 | 0.512 |

Censore: censoring rate; RMSTd: test based on RMSTd; ABS: ABS test; ABSP: ABS Permutation test.